\documentclass[doublecol]{epl2}

\usepackage{graphicx,color,amsmath,amssymb}
\newcommand{\halfsq}{\frac{1}{\sqrt{2}}}
\newcommand{\half}{\frac{1}{2}}

\title{Semi-counterfactual    Cryptography}
\author{Akshata  Shenoy H. \inst{1} \and
   R. Srikanth \inst{2,3} \and T.   Srinivas \inst{1}}

\institute{   \inst{1}   ECE  Dept,   IISc,   Bangalore  \\   \inst{2}
  Poornaprajna  Institute  of  Scientific Research,  Bangalore,  India \\
  \inst{3} Raman Research Institute, Bangalore, India 
}


\abstract{  In  counterfactual  quantum  key distribution  (QKD),  two
  remote parties  can securely share  random polarization-encoded bits
  through the blocking rather  than the transmission of particles.  We
  propose a semi-counterfactual QKD, i.e., one where the secret bit is
  shared, and also encoded, based on the blocking or non-blocking of a
  particle.  The  scheme is thus semi-counterfactual and  not based on
  polarization encoding.  As with other counterfactual schemes and the
  Goldenberg-Vaidman  protocol, but unlike  BB84, the  encoding states
  are orthogonal  and security arises  ultimately from single-particle
  non-locality.  Unlike any of  them, however, the secret bit generated
  is maximally indeterminate until the  joint action of Alice and Bob.
  We prove  the general security of  the protocol, and  study the most
  general photon-number-preserving incoherent attack in detail.  }

\begin{document}
\maketitle

\section{Introduction}

Quantum key distribution  (QKD) is a method allows  two parties (Alice
and Bob) to share a secret key, whose secrecy is protected by the laws
of   quantum   mechanics   (QM),    such   as   no-cloning   and   the
indistinguishability  of   non-orthogonal  states  \cite{qkdrmp}.   It
remains the  most advanced  application of quantum  information theory
experimentally  \cite{scarmp},  and   even  commercially.   Since  the
proposal of  the first QKD protocol \cite{bb84},  various paradigms of
QKD have been proposed  such as use of entanglement \cite{ekert,lc00},
orthogonal    states     \cite{gv,ABD+10},    two-way    communication
\cite{pingpong,lm05,    SPS+12},     secure    direct    communication
\cite{qsdc04,qkd11,cas12, EM11} and, most recently, counterfactual QKD
(CQKD)   \cite{noh09,salih},   which  is   based   on   the  idea   of
interaction-free  measurement \cite{elva}.   CQKD involves  secret key
sharing through blocking rather  than transmission of a particle.  The
Noh  protocol   \cite{noh09}  has  since  been   made  more  efficient
\cite{cqkdeff}        and       its        security       investigated
\cite{cqkdsecu1,cqkdsecu2,trojan}.   Recently,  it was  experimentally
implemented \cite{cqkdexp}.

In  this  work,  we describe  a  scheme  for  CQKD which,  unlike  the
counterfactual  protocols of  Refs. \cite{noh09,salih},  does  not use
polarization encoding.   To see why  this is interesting, it  is worth
noting  that  the  idea   of  information  transfer  without  particle
transmission   really   means   information   transfer   by   blocking
signals. This can  be done classically, too. A  `classical' version of
the Noh protocol would be: Alice  transmits a red or blue ball to Bob,
who randomly applies a  color-based blocking operation.  If he applies
a red-blocker,  and Alice  transmitted a red  ball, then he  holds 
back  the  ball,  else  he  transmits it  back.   Similarly,  for  the
blue-blocker.   Alice and  Bob share  a  bit, determined  by the  ball
color, whenever  Alice does  \textit{not} receive a  ball.  Naturally,
this  classical  protocol  is  completely  insecure  because  Eve  can
determine the color  of the ball during the  forward transmission, and
find out if Bob returns  the ball.  In the quantum mechanical version,
however, Bob's blocking can  spoil distructive interference on Alice's
side, leading  to a particle  detection far away from  Bob's blockade,
which is the  counterfactual element.  If Eve tries  to find the color
(read:  polarization), she commits  the ball  (read: particle)  to the
Alice or Bob path, potentially disrupting the path superposition state
in which the particle would  otherwise be prepared, which is the basis
of security.

Now  if both  bits are  to be  generated by  Bob's blocking  act, then
encoding  on  basis  of  some  internal degree  of  freedom  (such  as
polarization   or  spin)  would   be  needed.    If  now   we  require
counterfactuality only on one bit value, then one of the bits would be
generated by blocking and the other by transmission. But in that case,
the act  of blocking  or non-blocking itself  could be used  to encode
bits,  thereby  `freeing'  polarization  from the  duty.   Alice  thus
transmits a  ball to Bob,  and he blocks  or does not block  it. Alice
deduces the bit according to  whether she receives or does not receive
his trasnmission.  The second  case represents the counterfactual bit.
Our semi-counterfactual QKD protocol  is the secure quantum version of
this  classical  idea.  Apart  from  the  conceptual clarification  it
provides to  the counterfactual paradigm  in cryptography, it  has the
practical benefit of  allowing polarization to be used  to be used for
security against certain Trojan horse attacks.

This article  is structured as  follows. In the following  section, we
present the new protocol, and  compare and contrast it with some other
related protocols in the subsequent section.  Thereafter, the security
of the protocol is discussed, followed  by a discussion on the role of
the translated  no-cloning theorem in  the present protocol.   Next we
prove the security under a  general incoherent attack in the noiseless
scenario.  However,  in practice, noise is ubiquitous.   This is taken
into  consideration in  the subsequent  section, where  we  derive the
tolerable  error  rate  for  a restricted,  incoherent  photon-number-
preserving attacks. 

\section{A new scheme \label{sec:new}}

Light from the source $S$ hits the beamsplitter BS and splits into two
beams  one   along  Alice's  arm   $a$ ($A$) and  another  Bob's   arm  $b$ ($A$)
respectively.The quantum state after BS is
\begin{equation}
|\phi\rangle_{AB}  =  \sqrt{T}  |00\rangle_A|\psi\rangle_B  +  i  \sqrt{R}
|\psi\rangle_A|00\rangle_B,
\label{eq:state}
\end{equation}
where  $T$ and  $R$ represent  the coefficients  of  transmittance and
reflectance  of  the  BS  respectively  such  that $T  =  1-R  $,  and
$|00\rangle$ represents the vacuum state in the two polarization modes
$H$ and $V$, while $|\psi\rangle$  represents a single photon state of
arbitrary  polarization,  i.e.,  $|\psi\rangle  =  \alpha|10\rangle  +
\beta|01\rangle$  with  $|\alpha|^2  +  |\beta|^2  =  1$.   The  first
(second) ket refers to the transmitted (reflected) or Alice (Bob) arm.

Alice and Bob each possess a  switch SW, an absorber Abs and a Faraday
mirror FM.  Each of the  participants either applies the operation $F$
(reflect) or $A$ (absorb) depending  on the state of the switch, which
is randomly  on or off,  respectively.  Ideally, the two  SW's are
independently controlled  by a  quantum random number  generator.  The
three possibilties are: (1) Both SW's are on, so that Alice and Bob
reflect their beams which  causes an interference pattern at the
detector $D_1$ ($D_2$) with  probability $(T-R)^2$ ($4RT$); (2) One of
the SW's is on while the other is off.  This results in a detection at
$D_2$ with probability $RT$, due to the physical travel of light along
only  one of  the arms.   If Alice  (Bob) did  the blocking,  then the
probability  for  detection  at  $D_1$  is  $T^2$  ($R^2$),  with  the
probability for  absorption at the respective module  being $R$ ($T$).
This  is  the  mode (on-off  or  off-on)  in  which  a secret  bit  is
generated. By  convention, it is labelled  either 0 or  1 depending on
whether Alice  or Bob  applies the $A$ operation. The  protocol is
counterfactual with respect to Bob when secret bit 1 is generated; (3)
The last possibility is that both the SW's are in the off mode resulting in
Alice and  Bob applying  the operation $A$  and there is  no detection
both at $D_1$ and $D_2$.

\begin{table}[h]
\begin{tabular}{c|c}
\hline  Detection   &  Pattern  (Alice,  Bob)  \\   \hline  $D_1$  &
($(F,A),\frac{1}{4}$),   ($(A,F),  \frac{1}{4}$)   \\  \hline   $D_2$  &
($(F,F),1$), ($(F,A),\frac{1}{4}$),  ($(A, F),\frac{1}{4}$) \\ \hline  Null &
($(F, A),\frac{1}{2}$), ($(A, F),\frac{1}{2}$), ($(A, A),1$) \\ \hline
\end{tabular}
\caption{The allowed  pattern of Alice's  and Bob's action  along with
  the probability that a detector click was produced.}
\label{tab:error}
\end{table}

The counterfactual nature of the protocol comes from the fact that the
secret bit 1  is extracted by the interaction-free  detection of Bob's
$A$ setting.  A detection at $D_1$ is determined by the presence of an
absorber which  did not  scatter the particle.   When secret bit  0 is
generated, it is counterfactual with  respect to the internal arm $a$,
but the photon travels physically  along Bob's arm $b$. In this sense,
the protocol  is only semi-counterfactual with respect  to the exposed
arm.

The corresponding  patterns of operations and  outcomes are summarized
in Table  \ref{tab:error}.  The efficiency of the  protocol is defined
by    the   probability    $   P(D_1)    =    P(D_1|(F,A))P((F,A))   +
P(D_1|(A,F))P((A,F)) = \frac{1}{4}\frac{1}{4} + \frac{1}{4}\frac{1}{4}
= \frac{1}{8}$,  which is the same  as in the Noh  protocol.

\begin{figure}
\onefigure[width=8.5cm]{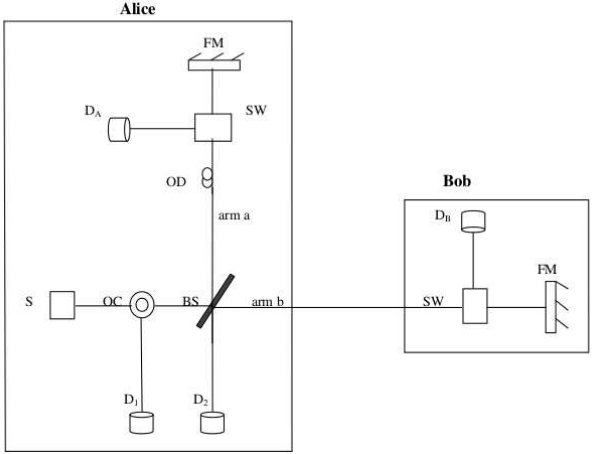}
\caption{Semi-counterfactual quantum cryptography:     The    experimental
  architecture    of   the    proposed   QKD    using   Michelson-type
  interferometer. Alice's station consisting of the source S initiates
  the protocol by sending  light pulses through the optical circulator
  OC to the  beamsplitter BS, which splits them  into beams along arms
  $a$ and  $b$. The optical delay  OD maintains the  phase between the
  arm by compensating for the  path difference in the two arms.  Light
  along arm a is subjected  to absorption or reflection by Alice based
  on  her  switch state.   Likewise  by Bob  along  arm  $b$ who  also
  possesses SW, Abs and FM.}
\label{fig:cf}
\end{figure}

The protocol  steps are as  follows: (I) $n$ photons  are sequentially
injected  from   the  left-hand  side  of  the   apparatus  in  Figure
(\ref{fig:cf}).   (II) Depending  on  the random  switch state,  Alice
applies $F$ or $A$  in the arm $a$ on each photon,  and so does Bob in
arm $b$. (III) On the $n$ outcome data collected, a fraction $f$ is randomly
selected, for  which Alice  and Bob both  announce their  settings and
outcome  information.  They  check  that the  observed  statistics  is
sufficiently close to Table \ref{tab:error}.  Two figures of merit are
the visibility of the interference fringes 
\begin{equation}
\mathcal{V}  \equiv  \frac{P(D_2|F,F)   -  P(D_1|F,F)}  {P(D_1|F,F)  +
  P(D_2|F,F)}.
\label{eq:visi0}
\end{equation}
and the error rate
\begin{equation}
e \equiv P(F,F|D_1) +  P(A,A|D_1),
\label{eq:e}
\end{equation}
which  estimates the fraction  of mismatched  secret bits.   Two other
figures of  merit are  estimates on $r$,  the rate of  multiple count,
which   may  be   due  to   dark  counts   or   certain  photon-number-
non-preserving attacks, and $\lambda$, transmission loss rate over the
channel.   (IV) If  $\mathcal{V}$  $(e)$ is  sufficiently  close to  1
$(0)$, then the  remaining approximately $(1-f)n/8$ bits corresponding
to $D_1$  detection are used for further  classical post-processing to
extract  a  smaller secure  key  via  key  reconciliation and  privacy
amplification.

If the  timings of  pulses where  not random, and  Eve knew  them, she
would probe Bob's  setting by entering a photon into  the path $b$ and
studying it upon return, using an Alice-like set-up.  As one method of
protection  against such  a  Trojan horse  attack \cite{trojan}  Alice
sends  her pulses  randomly. She  and Bob  later compare  that  in the
events where  Bob applied the operation $A$  \textit{and} detected the
photon, the  time $t_i$  when the photon  was sent by  Alice satisfies
$t_s =  t_r -  \tau$, where $t_r$  is the  time of Bob's  receipt, and
$\tau$ the  photon transit time.  As a simpler alternative,  Alice can
vary  the polarization  randomly between  the BB84  states $|V\rangle,
|H\rangle,\frac{1}{\sqrt{2}} (|V\rangle \pm  |H\rangle)$ and Bob's $A$
operation may  be accompanied  by a polarization  measurement.  Trojan
horse attacks by Eve to probe  Bob's settings can be detected by later
verifying the polarization in his detection events.

\section{Comparison and contrast with some other protocols \label{sec:comp}}

It is worth comparing and contrasting the present protocol with others
that employ  single photons: e.g., the  counterfactual protocols, BB84
\cite{bb84}  or  the Goldenberg-Vaidman  protocol (GV) \cite{gv}.  In  all
these cases,  except BB84 (which uses conjugate  coding), the encoding
is  with orthogonal states.   In particular,  the actions  $(A,F)$ and
$(F,A)$,  when  they  can  lead  to a  $D_1$  detection,  produce  the
orthogonal        states,       $|00\rangle_A|\psi\rangle_B$       and
$|\psi\rangle_A|00\rangle_B$,  respectively.    As  with  these  other
orthogonal-state-based  protocols,  single-particle non-locality,  and
restricting  the  observables that  Eve  can  access,  is the  key  to
security.

Our protocol,  however, differs from  the other orthogonal-state-based
protocols in  one respect worth noting. The  initial state transmitted
by  Alice is  the same  in all  cases.  Only  after Alice's  and Bob's
actions  is  the secret  bit  generated.  In  GV,  the  secret bit  is
deterministic,  while   in  the   Noh  protocol,  Alice   decides  the
polarization  beforehand, so  that if  a detection  is  accepted (with
probability = 1/8),  then the secret bit is  fixed by the polarization
chosen earlier. By  contrast, in the present protocol,  the secret bit
is decided only after Alice's and Bob's actions.
 
\section{Security \label{sec:secu}}

Intuitively, security  arises from single-particle  nonlocality: Eve's
attempt to determine  Bob's action will tend to  localize the particle
into  one or the  other arm,  forcing a  particle nature,  and thereby
disrupting  the  coherence between  the  two geographically  separated
wave packets in the two arms. This will be reflected in a reduction of
visibility, which can be  observed.  However, the protocol is two-way,
meaning  that in some  form (either  physically or  as a  vacuum), the
particle is  re-exposed after Bob responds,  and one needs  to be sure
that  Eve  cannot  exploit  the  difference between  the  ingoing  and
outcoming states.

When the  secret bit  generated is 1,  then Bob applied  the operation
$A$, and the photon never physically travelled on the exposed arm $b$.
This seems to make the generation  of this bit inherently safe from an
eavesdropper Eve monitoring the arm.  This is the intuition behind the
expectation that  counterfactuality helps security.   However, quantum
optically, non-travelling of  a physical particle entails the  travel of a
vacuum pulse,  with a non-trivial physical consequence.   Thus a claim
for security cannot be made without more detailed consideration.

\subsection{Translated no-cloning theorem \label{sec:trans}}

In the Noh  protocol, the reduced density operator  of the particle in
the    exposed    arm    during    the   forward    transmission    is
polarization-dependent,  and thus  correlated with  the  secret state.
However, the correlated states  are non-orthogonal, and the protocol's
security  is  attributed to  this  non-orthogonality.  In  particular,
Eve's  action would  be  to evolve  the  exposed wave  packet with  an
ancilla    prepared    in     state    $|m\rangle{_E}$    such    that
$U|s_j\rangle{_B}|m\rangle{_E}  = |s_j\rangle{_B}|m_j\rangle{_E}$ with
$j=1,2$.   However, unitarity demands  that $\langle  s_1|s_2\rangle =
{_B}\langle s_1|s_2\rangle{_B} {_E}\langle m_1|m_2\rangle_E$, implying
that for  cloning to work, either $|s_1\rangle$  and $|s_2\rangle$ are
orthogonal   or  $|m_1\rangle   =  |m_2\rangle$   and  Eve   gains  no
information. Thus Eve's any action that gains information will disturb
the   state    of   Alice's   and   Bob's    particle,   implying   an
information-vs-disturbance  trade-off. In  the  backward transmission,
both bit values are represented by the vacuum state.

In the present  protocol, the state of the particle  sent out by Alice
is  the  same  for  either bit,  namely  $\half|00\rangle\langle00|  +
\frac{1}{4}\left(|01\rangle\langle01|  + |10\rangle\langle10|\right)$.
But during the backward  transmission, the states corresponding to the
secret bits are  represented in the exposed arm  by $|00\rangle_B$ and
$\frac{1}{2    }\left({_B}|01\rangle\langle01|_B    +   {_B}|10\rangle
\langle10|_B\right)$, which are orthogonal.  The above reasoning would
imply that our protocol is insecure.   We will find below that this is
not the case, mainly because the encoding states are not orthogonal to
$|\phi\rangle_{AB}$,  and  that here  too  we  have  a situation  with
information-vs-disturbance trade-off.

\subsection{General incoherent attack\label{sec:gen}}

In the most general, coherent attack, Eve prepares $n$ ancillas in the
state $|\epsilon_1\rangle^{\otimes  n}$ and  another $n$ in  the state
$|\epsilon_2\rangle^{\otimes n}$.  She has each particle interact with
an  ancilla  prepared  in  state  $|\epsilon_1\rangle$  using  unitary
$\mathcal{U}^O$ in the onward leg, and then with one prepared in state
$|\epsilon_2\rangle$ using unitary  $\mathcal{U}^R$ in the return leg.
She waits until after Alice's announcement, to select a suitable basis
to jointly measure all her ancillary pairs.

We will focus on Eve's most general incoherent attack. Eve prepares an
ancilla in state $|0\rangle$,  and applies the operation (\ref{eq:op})
on the Bob-Eve system. We  will assume some fixed polarization in this
and the next subsection, and replace a double-ket with a single-ket to
represent Alice's and Bob's particles, for simplicity.
\begin{eqnarray}
|0\rangle_B|0\rangle_E  &\rightarrow& 
\alpha_{00}|0\rangle_B|\epsilon_{00}\rangle_E + \alpha_{0\perp}|0^\perp\rangle_B
|\epsilon_{0\perp}\rangle_E \nonumber \\
|1\rangle_B|0\rangle_E  &\rightarrow& 
\alpha_{10}|1\rangle_B|\epsilon_{10}\rangle_E + \alpha_{1\perp}|1^\perp\rangle_B
|\epsilon_{1\perp}\rangle_E,
\label{eq:op}
\end{eqnarray}
with $|\alpha_{00}|^2 +  |\alpha_{0\perp}|^2 = 1$ and $|\alpha_{10}|^2
+    |\alpha_{1\perp}|^2    =    1$,   and    $|0^\perp\rangle$    and
$|1^\perp\rangle$  are  states  orthogonal  to the  vacuum  state  and
$|1\rangle$,  respectively.  The  fixed  state  transmitted  over  the
channel and the ancilla evolves under (\ref{eq:op}) to

\begin{widetext}
\begin{equation}
\frac{1}{\sqrt{2}}(|0,1\rangle   +   |1,0\rangle)_{AB}   |0\rangle_E
\rightarrow \halfsq\left(|0\rangle\left[\alpha_{10}|1\rangle|\epsilon_{10}\rangle
 + \alpha_{1\perp}|1^\perp\rangle|\epsilon_{1\perp}\rangle\right] +
|1\rangle\left[\alpha_{00}|0\rangle|\epsilon_{00}\rangle
 + \alpha_{0\perp}|0^\perp\rangle|\epsilon_{0\perp}\rangle\right] 
  \right).
\label{eq:genataq}
\end{equation}
\end{widetext}
\begin{floatequation}
\mbox{\textit{see eq.~\eqref{eq:genataq}}}
\end{floatequation}
It suffices for our qualitative treatment to consider the state of the
particle in the event when Alice and Bob apply the operation $FA$, and
Bob registers  a detection.   It follows from  Eq.  (\ref{eq:genataq})
that the state of the particle is:
\begin{equation}
\halfsq\left(\alpha_{10}|0\rangle_A|0\rangle_B|\epsilon_{10}\rangle_E
 + \alpha_{0\perp}|1\rangle_A(\hat{b}|0^\perp\rangle)_B|\epsilon_{0\perp}\rangle_E
\right),
\label{eq:bobabsorb}
\end{equation}
where $\hat{b}$  is the  annihilation operator on  Bob's mode,  and we
have conservatively  set $|1^\perp\rangle \equiv  |0\rangle$, implying
that  there is  a  probability $|\alpha_{0\perp}|^2$  of  a double  or
multiple count, which can be detected by Alice in principle.

We note that this detection occurs irrespective of whether Eve follows
up with a further attack on the return leg, in particular, even if she
`unattacks', i.e., reverses the operation (\ref{eq:op}). The danger of
producing multiple counts can be avoided  by Eve if she uses a weaker
attack that conserves photon number. We study this in the remainder of
this section in a general setting.

\subsection{General photon-number-preserving incoherent attack\label{sec:num}}

The above general attack only indicates that the protocol is secure if
no noise is  detected. In practice noise will  be unavoidable, even in
the absence of an eavesdropper.  On the other hand, conservatively, we
must  assume  that  all noise  is  due  to  Eve, and  upper-bound  her
information on the secret key, from which the tolerable error rate can
be derived.  This task can be  difficult for a general model of attack
as above,  and we  adopt a simpler  one here.  On the onward  leg, Eve
prepares  an ancilla  in the  state  $|0\rangle_E$, and  in the  joint
system $BE$, applies the number-preserving operation
\begin{equation}
\mathcal{U} = |00\rangle_B\langle00|\otimes U_0 + 
   (|01\rangle_B\langle01| + |10\rangle_B\langle10|) \otimes U_1,
\label{eq:atakU}
\end{equation}
such that $\langle  0|U^\dag_1U_0|0\rangle \equiv \langle Y|N\rangle =
\cos(\theta)$. Eve's attack in the
onward leg followed produces the state:
\begin{equation}
\mathcal{U}|\phi\rangle_{AB}|0\rangle       =      \halfsq(|\psi\rangle
|00\rangle|N\rangle + |00\rangle |\psi\rangle|Y\rangle).
\label{eq:statevac0}
\end{equation}
The Alice-Bob action  $(F,F)$ leaves the $|\phi\rangle_{AB}$ unchanged.
In the case  of $(F,A)$, the resulting states  are $\halfsq |00\rangle
|00\rangle |Y\rangle$  or $\halfsq|\psi\rangle|00\rangle|N\rangle$, of
which  the  former implies  detection  by  Bob  and the  latter  leads
potentially to  a $D_1$ detection  for secret bit  1.  In the  case of
$(A,F)$,  the  resulting  states  are $\halfsq  |00\rangle  |00\rangle
|N\rangle$  or $\halfsq|00\rangle|\psi\rangle|Y\rangle$, of  which the
former implies detection by Alice  and the latter leads potentially to
a $D_1$ detection for secret bit 0.

Suppose  on the  return leg,  she `unattacks'  the same  system, i.e.,
applies $\mathcal{U}^\dag$, and measures  the ancilla using a suitable
positive operator-valued measure  (POVM).  After the ``unattack'', the
case  $FF$  results  in  $|\phi\rangle_{AB}|0\rangle$,  implying  null
disturbance and also no information  for Eve. In the case $(F,A)$, the
resulting  states are  $\halfsq|00\rangle|00\rangle |\tilde{Y}\rangle$
(where    $|\tilde{Y}\rangle    \equiv    U^\dag_0U_1|0\rangle$)    or
$|\psi\rangle|00\rangle|0\rangle$.  In the case $(A,F)$, the resulting
states     are      $\halfsq|00\rangle|00\rangle     |0\rangle$     or
$\halfsq|00\rangle|\psi\rangle|0\rangle$.   Thus  a  detection of  the
probe  is in  the final  state  $|0\rangle$ whenever  the particle  is
available  for a  $D_1$  detection,making Eve  equally  unable to  say
whether  a  0  or  1  bit secret  was  generated.   Allowing  $\langle
\tilde{Y}|0\rangle = 0$, Eve knows that Bob applied $A$, but these are
precisely those events where Bob  has a detection, making the particle
unavailable for the secret-bit generation.

More  generally, Eve may  wish to  use a  general $\mathcal{U}^\prime$
rather than $\mathcal{U}^\dag$  in the return leg. Then,  if Alice and
Bob      applied      $(F,F)$,       the      final      state      is
$\halfsq(|\psi\rangle|00\rangle|N^\prime\rangle                       +
|00\rangle|\psi\rangle|Y^\prime\rangle$,   where  $|N^\prime\rangle  =
U_0^\prime   U_0|0\rangle$    and   $|Y^\prime\rangle   =   U_1^\prime
U_1|0\rangle$.   In  the  case   $(F,A)$,  the  resulting  states  are
$\halfsq|00\rangle|00\rangle      |\tilde{Y}^\prime\rangle$     (where
$|\tilde{Y}^\prime\rangle     \equiv    U^\prime_0U_1|0\rangle$)    or
$|\psi\rangle|00\rangle|N^\prime\rangle$.   In the  case  $(A,F)$, the
resulting states are $\halfsq|00\rangle|00\rangle |N^\prime\rangle$ or
$\halfsq|00\rangle|\psi\rangle|Y^\prime\rangle$.   In   the  cases  of
relevance, namely the  cases $(F,F)$ and the subset  of events leading
to $D_1$  detection following applications of $(F,A)$  or $(A,F)$, the
state  of   the  probe  is  obtained  by   replacing  $|N\rangle$  and
$|Y\rangle$ with $|N^\prime\rangle$ and $|Y^\prime\rangle$ in relation
to the situation where no attack takes place in the return leg.

Thus, the  most general incoherent-number-preserving attack  that Eve
can  launch would  be to  use the  above onward  leg attack,  and then
measure her probe $E$ after  Alice's announcement.  We will assume the
worst-case  scenario   where  Eve   has  complete  knowledge   of  the
transmission  schedule between  Alice  and Bob.   Thus  she times  her
attack  to happen  just  when the  particle  is about  to enter  Bob's
station, and completes it after Alice's announcement of $D_1$ events.

It  follows from  the above  that  in such  events, Eve  would try  to
determine the secret bit by measuring her probe. She can optimally try
to distinguish  between $|N\rangle$ and  $|Y\rangle$, which correspond
to secret bits 0 and 1, respectively, using the following optimal POVM
\cite{nc00}:
\begin{eqnarray}
M_N &=& \frac{1}{1+|\langle N|Y\rangle|}(1 - |Y\rangle\langle Y|), \nonumber \\
M_Y &=& \frac{1}{1+|\langle N|Y\rangle|}(1 - |N\rangle\langle N|), \nonumber \\
M_{\rm inconcl} &=& 1 - M_N - M_Y,
\label{eq:povm}
\end{eqnarray}
where   outcome   $M_N$   ($M_Y$)  indicates   deterministic   outcome
$|N\rangle$   ($|Y\rangle$)  and   $M_{\rm  inconcl}$   represents  an
inconclusive  outcome. It  follows from  Eq. (\ref{eq:povm})  that the
probability  of  a conclusive  read-out  is  $  P_c \equiv  1  -
|\langle N|Y\rangle| = 1 - \cos(\theta) $.

Eve's information
\begin{equation}
I_{BE} = P_c = 1-\cos(\theta) = I_{AE} \equiv I_E,
\label{eq:I_E}
\end{equation}
the latter two equalities following from the symmetric nature of Eve's
information after Alice's announcement. In  the given
direction of polarization of the  photon, Alice's beam splitter may be
represented as:
\begin{eqnarray}
d_1^\dag &=& \frac{1}{\sqrt{2}}(a^\dag + i b^\dag) \nonumber \\
d_2^\dag &=& \frac{1}{\sqrt{2}}(a^\dag - i b^\dag),
\end{eqnarray}
where $a^\dag,  b^\dag$ are the  creation operators for the  modes $A,
B$,  respectively,  and $d_1$  and  $d_2$  are annihilation  operators
corresponding to  detections at $D_1$ and  $D_2$, respectively. Hence,
the state $|\phi\rangle_{AB}$ evolves to
\begin{equation}
|\phi^\prime\rangle             =             \frac{1}{\sqrt{2}}\left(
d_1^\dag|0\rangle_{AB}|{+}\rangle_E                                   +
id_2^\dag|0\rangle_{AB}|{-}\rangle_E\right),
\label{eq:prob}
\end{equation}
where   $|\pm\rangle_E   \equiv   \frac{1}{\sqrt{2}}(|N\rangle_E   \pm
|Y\rangle_E)$.  From this, it follows that
\begin{eqnarray}
\textrm{Prob}(D_2|FF)
&=& \frac{1}{2}|||N\rangle_E + |Y\rangle_E||^2 \nonumber \\
&=& \frac{1}{2}(1 + \cos(\theta)).
\label{proba}
\end{eqnarray}
We thus find that the visibility (\ref{eq:visi0}), conditioned on both
applying $F$, falls from 1 to
\begin{equation}
\mathcal{V} = \cos(\theta).
\label{eq:visi}
\end{equation}

For any other of the three pairs of operations by Alice and Bob (e.g.,
($A,F$)), an entanglement  of the form (\ref{eq:statevac0}) does not
arise, and  Eve's attack does not produce a deviation from
the pattern in Table \ref{tab:error}.  

Combining   Eqs.  (\ref{eq:visi})  and   (\ref{eq:I_E}),  we   find  a
complementarity relation
\begin{equation}
\mathcal{V} + I_E = 1,
\label{eq:complem}
\end{equation}
for the  visibility during the  $(F,F)$ instances and  the information
Eve gains during the $D_1$ outcomes. If the error rate is $e$, we have
Bob's information to  be $I_{AB} = 1 - h(e)$, where  $h$ is the binary
entropy.  Assuming  conservatively that there  is no other  noise than
that  due to  Eve, and  that she  employs the  above attack,  the only
changes    to   Table   (\ref{tab:error})    is   the    addition   of
$(F,F,\half(1-\cos(\theta))$ in  the $D_1$ row and  modifying the $FF$
entry  in the  $D_2$ row  to $(FF,\half(1+\cos(\theta))$.   Thus error
rate $e$ in Eq.  (\ref{eq:e}) becomes, by Bayesian rule,
\begin{eqnarray}
e   &=&  P(F,F|D_1)  =  \frac{P(D_1|F,F)P(F,F)}{P(D_1)}  \nonumber \\
&=& \frac{1-\cos(\theta)}{2-\cos(\theta)}.
\label{eq:ee}
\end{eqnarray}
This is an estimate of the error in key, since $D_1$ events arising
from an operation $(F.F)$ leads to a mismatch in the private copies
of the key.
The condition for positive key rate in the protocol is \cite{ck78}
\begin{equation}
K = I_{AB} - \min\{I_{AE}, I_{BE}\} > 0,
\label{eq:poskey}
\end{equation}
where $K$ are the secret bits that can be distilled after Alice and Bob
perform key  reconciliation and privacy  amplification.  The condition
for security  in our  protocol becomes, from  Eqs.  (\ref{eq:complem})
and (\ref{eq:poskey}),
\begin{equation}
h\left(\frac{1-\cos\theta}{2-\cos\theta}\right) < \cos\theta.
\label{eq:secu}
\end{equation}
or $\theta \lesssim 0.745$ rad, which, in view of 
Eq. (\ref{eq:e}), implies $e \lesssim 20.9\%$.

\section{Discussion and conclusions \label{sec:discu}}

We have presented a protocol for QKD which is counterfactual on one of
the  generated secret  bits in  that the  encoding corresponds  to the
blocking or the  non-blocking by Bob of a  transmitted particle.  This
is  motivated   by  observing   that  the  information   transfer  via
non-transmission  of a  particle in  the return  leg is  possible even
classically \cite{Gis13},  as we noted  earlier. Thus, this  cannot be
the basis for security.  What this  method does achieve is to make the
information transfer  counterfactual, whereby Alice is  alerted to his
blocking action by  the photon detection at a  detector where it would
otherwise  be absent  (the  fact that  Bob's  blockade influences  the
distant detector is the counterfactual element here \cite{Vai13}).  By
contrast, in a classically equivalent protocol, Alice would be alerted
to his blocking by the non-return of the particle.

In the Noh  protocol, Eve's attempt to eavesdrop  the channel can lead
to a  blockade-like effect depending  on the sharpness  \cite{BL84} of
her  eavesdropping operation, thereby  potentially leading  to Alice's
counterfactual detection of  the photon even when Bob  does not block.
This situation is  not unlike in a conventional  QKD protocol, such as
BB84,  where   Eve  disturbs  the  particle  when   her  operation  is
inconsistent  with Bob's.  The fact that  counterfactuality does  not seem  to
offer any extra cryptographic advantage (though it gives a new form of
communication  in  quantum  cryptography)   forms  the  basis  of  our
departure from the Noh protocol. 

Like the Noh, Pingpong, LM05 and Deng-Long \cite{qkd04} protocols, our
scheme may  be thought of as  a two-way QKD protocol.   In common with
these,  a  practical  implementation  of our  protocol  faces  certain
difficulties  because the  photon has  to travel  twice  the distance,
which can limit range and  key rate. Moreover, the two interferometric
arms exist  in different environments  (an external fiber link  to Bob
and  a spool  at Alice's  station), both  requiring  stabilization. We
believe that  these issues pose technological challenges,  that can in
principle be surmounted.

To protect the Noh  protocol against Trojan-horse attack, one requires
time-randomization of  the sequence, for  which the difficult  task of
rapid pulse referencing is required. With time randomization, the rate
at which  the key is generated  is not constant,  which is undesirable
for  network  integration.   By   contrast,  because  we  do  not  use
polarization  encoding, we  have used  it for  detecting  Trojan horse
attacks, as  discussed.  Thus time  randomization is not  required and
pulse  referencing is  less stringent  in our  protocol.  Furthermore,
because  encoding is  not polarization  dependent,  polarization drift
poses a lesser problem in stabilizing interferometer. 

Our protocol differs  from each of the above  two-way protocols in one
or more  key aspects.  Our protocol  has Alice sending  a fixed state,
unlike  in the  Noh  and LM  protocols.   It involves  single-particle
nonlocality unlike in the LM protocol; finally, unlike in the Pingpong
protocol, it does not  involve two particle entanglement.  A practical
advantage  of not  using polarization  encoding is  that it  makes the
protocol secure against a kind  of Trojan horse attack.

We have  assumed zero transmission  losses, so that every  particle is
accounted for  by Alice's  or Bob's detectors.  Thus, a  direction for
generalizing  our  work  is  to  allow for  lossy  channels.   Another
direction  is to study  how much  a more  general incoherent  and even
coherent attack, helps Eve.

\bibliographystyle{eplbib}
\bibliography{axtaip}

\end{document}